# Abstractions for AI-Based User Interfaces and Systems

ALEX RENDA, Cornell University
HARRISON GOLDSTEIN, Cornell University
SARAH BIRD, Facebook
CHRIS QUIRK, Microsoft Research
ADRIAN SAMPSON, Cornell University

Novel user interfaces based on artificial intelligence, such as natural-language agents, present new categories of engineering challenges. These systems need to cope with uncertainty and ambiguity, interface with machine learning algorithms, and compose information from multiple users to make decisions. We propose to treat these challenges as language-design problems. We describe three programming language abstractions for three core problems in intelligent system design. First, *hypothetical worlds* support nondeterministic search over spaces of alternative actions. Second, a *feature type system* abstracts the interaction between applications and learning algorithms. Finally, constructs for *collaborative execution* extend hypothetical worlds across multiple machines while controlling access to private data. We envision these features as first steps toward a complete language for implementing AI-based interfaces and applications.

## 1 INTRODUCTION

Rapid progress in machine learning has sparked a stampede toward new kinds of user interfaces based on natural interaction. Voice-directed assistants from Apple, Microsoft, Amazon, and Google combine speech recognition, natural language processing, and a vast array of backend capabilities to create the illusion of an intelligent human assistant. The emergence of these assistants coincides with a ballooning interest in chatbots and conversational user interfaces for tasks from customer support to IT system administration and medical diagnostics [10–12].

Excitement over AI-based user interfaces, however, has run ahead of the engineering tools that we need to implement them. Engineers complain of a new category of pitfalls that arise from building systems around machine learning [15, 16]. In this paper, we argue that the system design challenges in intelligent user interfaces deserve new programming language abstractions. We identify three core challenges. First, AI-based interactions lead to ambiguity. Unlike with button presses and menu selections, applications need to provide domain-specific evidence to help resolve user intent and choose the best action among a space of alternatives. Second, connections between applications and machine learning systems consist of verbose, error-prone boilerplate code. Finally, AI-based systems need to incorporate evidence from multiple users and compute on multiple machines while respecting the privacy of sensitive data.

We propose to address these engineering challenges as language-design problems. Our new abstractions are founded on the insight that applications should neither completely decouple their core logic from AI-based interface components nor arbitrarily intertwine the two concepts. A natural language interface, for example, should not be implemented as a shim that "translates" from spoken utterances into domain-specific commands: the meaning of utterances can depend on the application's state. At the same time, developers should not need to pervasively modify an application to add natural-language capabilities.

As a running example, consider a natural-language agent that can help schedule calendar meetings.[1] A decoupled implementation might train a model to deduce times, places, and other preferences from a natural-language request and then hand these specific parameters over to a calendar service API. Ideally, though, such a system should merge the natural-language suggestions with hard constraints from the domain, such as free/busy information from each attendee; with soft priors such as the user's historically preferred times of day; and even with complex operations on the domain data such as rescheduling of conflicting meeting requests. Exploring the space of alternatives requires careful bookkeeping and even coordination between multiple systems: all participants in the meeting should provide availability information but wait for a consensus decision before committing to specific time. Furthermore, each interaction with the user is an opportunity to explore and refine a model of the user's preferences. We envision programming languages and runtime systems that help programmers orchestrate the composition of domain data and AI-based decision making.

---
[1]Already there are email-driven agents for this task. For example: https://x.ai, https://calendar.help, and https://claralabs.com

At the same time, however, total integration between machine-learning models and application code is also not the right solution. Programmers should not, for example, need to rewrite a calendar application in a probabilistic programming language such as Church [8] just to add natural-language interface. In contrast to the literature on probabilistic programming, where the goal is to express sophisticated probabilistic models in a constrained programming model, we propose to embed a straightforward representation for uncertainty in a general-purpose language. First, a huge body of application code need not be probabilistic: API calls to query free/busy information or to update calendar items, for example, do not require probabilistic methods. Second, a simple uncertainty representation based on $n$-best lists often suffices where more sophisticated approaches come at conceptual and computational costs. For example, in spoken language understanding, most systems represent speech recognition results as a simple confusion network ($n$-best list of words at each position). Although a lattice would capture ambiguity more faithfully, confusion networks achieve great speedups with little loss in accuracy [9]. Third, maintaining a high-level interface to uncertainty allows the developer to remain agnostic to the details of the machine learning algorithms. Our proposed abstractions help the programmer deal with the inputs to machine learning algorithms (i.e., features) and the outputs from models (i.e., predictions), but not with the task of designing a bespoke ML model—which is best left to probabilistic programming languages or model toolkits such as TensorFlow [1].

In this paper, the idea is not to design new machine learning models or natural language processing capabilities; instead, we aim to make ML and AI techniques easier to apply in real systems.

This paper describes three language constructs in a new programming language, Opal, that abstract three core concepts in developing AI-enabled applications:

- **Hypothetical worlds** (Section 2), which let applications explore many alternatives, such as interpretations of ambiguous input.
- A **feature abstraction** (Section 3), which lets applications communicate domain and context information to machine learning systems without tying them to a specific algorithm or implementation.
- **Collaborative execution** of hypothetical computations (Section 4), which supports privacy-preserving interactions across users and systems.

These initial abstractions provide a conceptual framework for engineering in the presence of uncertainty of user intentions and goals. Section 5 discusses future work, open questions and opportunities in programming languages and systems for AI applications.

## 2 HYPOTHETICAL WORLDS

AI-based user interfaces often need to help users choose a best action among a space of alternatives. The action could be scheduling a meeting, booking a flight, or sending a message. The common thread is that the best alternative depends on how the world *would* look if the action were taken: in other words, the fitness of a potential action depends on its effect on the world.

We introduce *hypothetical worlds*, a language construct that expresses nondeterministic choice, to search for a best action among alternatives. Ambiguous user input, for example, can induce a range of potential actions. The idea is to let programs experiment with multiple possible interpretations of incomplete evidence before deciding which interpretation is most likely correct. Programmers can write code to try different hypotheses using a natural style, as if the code were interacting with the real world, but only commit to changes based on the outcome of the potential changes. There are two fundamental operations: a new **hyp** statement runs a block of code in an isolated, hypothetical context and returns a *world* value; and a `commit` operation applies the effects from a world value to the currently-executing world.

In our running calendar example, say the user asks to schedule a new lunch meeting at noon but not on which day. The application needs to find the day where the meeting fits in best with existing schedule constraints. A traditional implementation would need to consider potential consequences such as canceling existing meetings and reducing each day's free time for working. With hypothetical worlds, the program can *pretend* to make a proposed modification on each day and then decide which to commit:

```
for (day in weekdays) {
  world = hyp {
    calendar.add(event, day);
    if (!constraints_violated(calendar)) { break };
  };
}
world.commit();
```

The `constraints_violated` function here can inspect the state of the hypothetical `calendar` in each world without knowing that its state is hypothetical. The language runtime buffers all updates to `calendar` inside each **hyp** block, and the `world.commit()` operation releases the buffered effects to update the user's real calendar.

The **hyp** are inspired by fork–join models for parallel programming [5] and consistency models for distributed databases [6]. To represent "forkable" values like `calendar` above, our prototype implementation uses in-memory persistent data structures [7] that support efficient snapshots, rollback, and merging. As in other systems with partially-ordered access to shared state, conflicting updates are possible; the application manually resolves conflicts in our prototype.

*Distributions and unbounded search.* The **hyp** primitive by itself expresses one hypothetical action at a time. To search large spaces of alternatives, we add a higher-level **search** construct that builds on **hyp** to fork an unbounded number of hypothetical worlds simultaneously. To specify a search space, the program uses a *distribution* value, which is a probability-weighted set. The **search** construct produces a new distribution on world values. For example, a calendar application might explore the possible start times for a new meeting:

```
worlds = search start in timerange(today.begin, today.end) {
  calendar.add(event, start);
  viols = constraint_violations(calendar);
}
worlds.minimize(w => w.viols).commit();
```

Here, the `timerange` function produces a distribution over times. In our language, distributions can be implemented as explicit sets or as sampling functions—for example, `timerange` may work by randomly sampling a minute from the given range. Then, the **search** construct produces a distribution value that can sample *world values* in which the event is added at different times. Finally, a standard-library `minimize` function samples from the world distribution to select a world with a minimal violation count. It takes a function that extracts a numeric score from the world to guide the sampling procedure.

The language's first-class support for probability distributions is inspired by work in probabilistic programming [8]. The **search** construct can build up larger sampling functions from smaller components, as in the Uncertain<T> language [3]. As in probabilistic programming languages, the minimize function uses a general sampling procedure such as a Markov chain Monte Carlo (MCMC) algorithm to search among possible executions. By coupling probabilistic programming techniques with hypothetical execution, our language lets applications use statistical search *in situ*, without switching to a fully probabilistic language.

*Composition and recursive search.* Hypothetical worlds compose to build up complex, nested search problems. Calling commit in a hypothetical world merges a child world's state into its parent. For example, to fit a new meeting into the user's calendar, a program can recursively reschedule any conflicting meetings:

```
function schedule(event) {
  worlds = search start in timerange() {
    calendar.add(event, start);
    conflicts = find_conflicts(calendar, event);
    for (conf in conflicts) {
      calendar.delete(conf);
      schedule(event);
    }
  }
  worlds.minimize(w => w.conflicts.length).commit();
}
```

This procedure removes each conflict from the calendar and then calls itself—at another level of hypothetical execution—to restore it. The recursive calls create a hierarchical search space that minimizes the number of event movements at each level.

*Iterative refinement.* In interfaces based on natural language, such as voice assistants and chatbots, interactions rarely occur in a single shot. Agents need to use a multi-turn, conversational structure where the user iteratively refines their intent to guide the assistant toward a desired outcome. The application needs to efficiently support interactions with this structure:

|  | *General structure* | *Calendar example* |
| --- | --- | --- |
| User: | Please take an action with parameters $P$. | Can you book a meeting with Jane and Alex? |
| Assistant: | OK, I found a few alternatives: actions $A_1$ and $A_2$ both satisfy $P$. | You are both free on next Monday at 1pm, or Tuesday at 9am. |
| User: | None of those work. How about with parameters $P'$, which slightly refine $P$? | That's too late. Could I just meet with Jane instead? |
| Assistant: | I found a new potential action $A_3$ that satisfies $P'$. | You are both free tomorow at 2pm. Should I schedule it? |
| User: | OK! Take action $A_3$. | Yes, book a 2pm meeting, thanks. |

A naive implementation would re-run each query from scratch using every refined set of constraints. Instead, we design hypothetical search to be *incremental* to let the system efficiently respond to slightly modified queries. The design is inspired by work on language support for incremental updates [2]. The runtime system memoizes the set of hypothetical world values created in each **search** invocation and tracks their dependencies on data from their parent worlds. When execution reaches the same `search` a second time, it opportunistically reuses hypothetical worlds whose inputs have not changes.

*Integrating with external services.* Because hypothetical execution isolates effects, it creates a natural point for controlling communication with the outside world. In our prototype, integration with external services—a CalDAV web service for storing calendars, for example—works by mapping Opal's persistent data structures to API calls. Communication occurs only outside of all **hyp** blocks: i.e., when execution is *not* hypothetical.

When the program makes hypothetical updates to a service-backed `calendar` data structure, for example, the runtime system buffers them in memory, as usual. When the program uses `commit` to merge the updates with the top-level "real" world, the runtime translates the new updates into API calls to modify the calendar on the server. This way, **hyp** blocks act as a concurrency control mechanism at the interface between Opal programs and legacy systems.

## 3 FIRST-CLASS FEATURES FOR LEARNING

Hypothetical worlds let applications express ambiguity, but to make decisions about *how* to resolve it, programs need to use machine learning. Typically, programmers need to write boilerplate interface code to marshal domain-specific values into a format that a general machine learning algorithm can consume. This *ad hoc* interaction with ML is error prone and verbose, and it complicates the implementation of more sophisticated interactions with ML engines. Furthermore, this bespoke interface code often ties applications to a specific machine learning algorithm or implementation.

We propose a new set of abstractions for improved clarity and safety in *feature extraction*. Here, a feature is a numeric or categorical value that describes aspects of data and context in the domain model. Programs extract collections of feature values from their internal data representation and provide *feature vectors* to machine learning algorithms for classification. While deriving and managing features can seem like a peripheral concern—all the "real work" happens in the machine learning algorithm itself—applications spend significant complexity on this part of their operation. Furthermore, the accuracy of a machine-learned system is equally if not more dependent on feature engineering than the particular learning algorithm.

### 3.1 Requirements: Feature Forms and IDs

Opal's feature abstractions act as an interface between the domain objects that provide context for decisions and datasets that are used as inputs to ML toolkits. Opal's feature interface works as a common intermediate language: domain objects can be translated into feature vectors, and feature vectors can be transformed into whatever format is necessary for the learning algorithm. Our interface must be powerful enough to make feature extraction easy, but also narrow enough to make it straightforward to incorporate new ML toolkits and data formats.

We categorize features into three *forms:*

- **Numeric** features are floating-point numbers that represent counts, distances, and so on.
- **Bounded** categorical features hold a data type that permits equality comparisons where values are drawn from a finite domain. A bounded feature includes a *bound*, which is a set of possible values. For example, the bound for a word feature might be a dictionary.
- **Unbounded** Finally, unbounded features are also categorical, but they do not require that there be a fixes set of categories. A URL, for example, is an unbounded feature.

A traditional approach to managing features associates an ID with every feature value extracted from the data. For example, a non-Opal system that extracts word-level features from text might use produce a string label for each word frequency value:

```
let out = [];
```

```
for word in tokenize(doc) {
  let freq = count(word, doc);  // Count the occurrences of word in doc.
  out.push("freq_" + word, freq);
}
```

The string `"freq_foo"` identifies the frequency feature for the word *foo*. ML toolkits rely on feature IDs to distinguish the entries in a feature vector: for example, a document classifier might compute cosine similarity between two documents by comparing the frequency features they have in common.

Using strings as feature IDs is convenient, but it is both inefficient and error prone. To construct each new feature, a system needs to use string concatenation and conversion operations to build up IDs. Subtle bugs can occur when the strings involved are mistyped even slightly: for example, searching for a feature with ID `"freq-foo"` instead of `"freq_foo"` will silently fail under this model. We propose a structured representation of feature IDs to avoid these pitfalls.

### 3.2 Safe Features with Dependent Types

We first describe a type-based abstraction for feature and feature IDs. The key idea is to represent feature IDs at the type level to statically guarantee that they are used safely. The approach builds on *dependent types* to express the interplay between types and values. While the design in this section captures the essence of feature manipulation, its reliance on a sophisticated type system limits its practicality; in the next section, we describe an encoding of the same ideas into a mainstream language without dependent types.

We first encode the three kinds of features—numeric, bounded, and unbounded—as a type. Using the syntax of Idris [4], a Haskell-like dependently typed programming language, we define a `Schema` data type:

```
data Schema : Type where
  N : FeatId -> Schema
  B : FeatId -> {a : Type} -> List a -> Schema
  U : FeatId -> Type -> Schema
```

`N`, `B`, and `U` represent *numeric*, *bounded*, and *unbounded* respectively. The `N` (numeric) constructor simply wraps a feature ID; the `U` (unbounded) constructor takes a type as a parameter to indicate the underlying feature value type; and the `B` (bounded) constructor keeps a copy of the dictionary that bounds the feature. Here, the type for feature IDs, `FeatId`, is not constrained: the application can use any type it likes, including strings or richer values that include domain-specific context. The static guarantees that the type system offers do not depend on the feature ID type.

We can now define the type of a feature:

```
data Feature : Schema -> Type where
  Numeric : (id : FeatId) -> Number -> Feature (N id)
  Bounded : Eq a =>
            (id : FeatId) ->
            (d : List a) ->
            (x : a) ->
            {auto prf : Elem x d} ->
            Feature (B id d)
  Unbounded : Eq a => (id : FeatId) -> a -> Feature (U id a)
```

The `Numeric` and `Unbounded` constructors are fairly straightforward, but the `Bounded` constructor relies heavily on dependent typing. Most notably, `{auto prf : Elem x d}` says that the `Bounded` constructor takes, as an implicit argument, a proof that the feature data is actually contained in the dictionary. This proof is easily automated by the type checker, so the result is a guarantee that a bounded feature is always constructed from valid data.

Finally, we define a list-like type for feature vectors:

```
data FeatureVector : List Schema -> Type where
  Nil : FeatureVector []
  (::) : Feature s ->
         FeatureVector slst ->
         FeatureVector (s :: slst)
```

With dependent types, the type of a feature vector includes the types for all the features that the vector contains. This way, the program can assert that each feature vector contains the set of features that the programmer expects. For example, the type system can statically rule out comparisons between two feature vectors when one is missing features that the other includes.

### 3.3 A Practical Realization

While dependently types can precisely capture the spirit of safe feature vector manipulation, they are not yet practical in mainstream programming languages. We propose a practical implementation for languages with standard object-oriented type systems: specifically, we target TypeScript [13]. Although our implementation of this feature abstraction lacks some of the static power, it accomplishes many of the same goals. In particular, it keeps the same interface design that makes translation to and from Opal features easy.

To express feature types, we define one class for each feature form. For example, the Numeric class implements a Feature interface and includes a data field with the TypeScript number type:

```
class Numeric implements Feature {
    form: "n";
    id: FeatId;
    data: number;
    ...
}
```

The form field uses a TypeScript *literal type* to tag the feature as a numeric feature. The tagging system lets us write the general Feature type as a union of the three feature classes:

```
type Feature = Numeric | Bounded<any> | Unbounded<any>
```

Then, code can use the tag-checking idiom in TypeScript to discriminate between the three cases.

In this design, feature vectors are implemented using a simple wrapper around an array of Features.

*Generators.* Whereas dependent types statically rule out unsafe uses of feature IDs, our practical implementation achieves a similar effect by restricting the way that code can construct features. In each of type-level feature classes, we use a curried generator static method. The one for Numeric is:

```
static generator(id: FeatId): (d: number) => Numeric {
    return (data) => {
        let f = new Numeric();
        f.form = "n"; f.id = id; f.data = data;
        return f;
    };
}
```

Instead of constructing a numeric feature directly, the programmer must use a feature ID to create a generator function. Then, all examples of that feature are "stamped out" with the same machinery. Reusing a single generator function to create features creates a connection between features from the same "template" that resembles the type-level constraints in the dependently typed version. Features from the same generator are guaranteed to have the same id value, so code can test for matching features using fast pointer equality checks.

*Feature Pattern Matching.* To make the feature library useful, we need to be able to define functions that operate on features and feature vectors. In the dependently typed version, we could statically enforce safe operations on features; in our practical implementation, we add constructs to help dynamically pattern-match on features.

The most basic utility, `matchFeature`, lets code use pattern matching on a single feature value. Its type is:

matchFeature : (Numeric $\to a$) $\to$ (Bounded $b \to a$) $\to$ (Unbounded $c \to a$) $\to$ (Feature $\to a$)

where the three function arguments play the roles of the branches in the pattern match. For example, consider a function `densify` that takes a feature and represents it as an array of numbers. This construct is useful for transforming sparse feature vectors into a dense representation for machine-learning toolkits that do not support sparse representations. Using `matchFeature`, the function can dispatch on the feature's form:

```
let densify = matchFeature<number[]>(
    (n: Numeric) => [n.data],
    <T>(b: Bounded<T>) => b.dict.map(x => (x === b.data) ? 1 : 0),
    <T>(u: Unbounded<T>) => { throw new Error(); },
);
```

(Because unbounded categorical features do not have a dense representation, that case throws an error.) The implementation of `matchFeature` uses `form` tags to dispatch to the branches.

We also define `matchTwoFeature` to match identically-tagged pairs of features. This combinator abstracts out not only dynamic dispatch but also a compatibility check: the two features must have the same ID to be processed together. The Opal standard library defines an absolute difference function using `matchTwoFeature`:

```
let absDiff = matchTwoFeature<number>(
    (n1: Numeric, n2: Numeric) => Math.abs(n1.data - n2.data),
    <T>(b1: Bounded<T>, b2: Bounded<T>) => (b1.data === b2.data) ? 0 : 2,
    <T>(u1: Unbounded<T>, u2: Unbounded<T>) => (u1.data === u2.data) ? 0 : 2,
);
```

To define operations on entire feature vectors or pairs of vectors, the program can map these kinds of functions over the feature arrays. With these transformations, programs can define transformations from data represented as Opal features into datasets that can be consumed by a specific ML toolkit.

## 4 COLLABORATIVE EXECUTION

Intelligent user interfaces often involve collaboration between multiple users. Calendar tasks involve coordination between multiple attendees; medical diagnostics require privacy-aware coordination between parties. To support collaboration, applications need to aggregate data from different users, possibly stored in different physical locations with different privacy constraints.

We extend Opal to let programs specify and enforce *disclosure policies* that control how users' private information is exchanged with other systems. There are two main pieces in the language extension: a *placement* construct, **at**, that controls where code runs, and an *access* construct, **with**, that controls data ownership. Together, the two language constructs limit the points where data owned by one user is disclosed to a machine owned by a different user or service. Then, the application can define custom policies to interpose on these disclosure points: they can ask the user for explicit permission every time, for example, or predict the risk of disclosure to avoid overwhelming the user. The design takes inspiration from existing work on language-based security [14, 17].

Consider a distributed calendaring application, where Alice wants to schedule a meeting with Bob, but neither wants to disclose their entire personal calendar to the other. However, both trust a central data center to handle the scheduling problem. Alice can run a **hyp** block that uses

```
world = hyp {
  at DataCenter {
    with Bob {
      time = find_available_time(Alice.calendar, Bob.calendar);
      Alice.calendar.add(event, time);
      Bob.calendar.add(event, time);
      fitness = Alice.fitness() + Bob.fitness();
    }
  }
};
if (world.fitness > threshold) { world.commit(); }
```

Listing 1. Alice requests that a central server node schedule a meeting using private data from both Alice and Bob.

**at** DataCenter to place the computation on the data center and **with** Bob to request access to Bob's data. Listing 1 shows an example in Opal. The **at** block runs code physically in the data center, and the **with** block allows it to access data from Bob. (Because Alice initiates the computation, her data is implicitly available.) Because the top-level code accesses the hypothetical fitness value, the **at** construct sends the value back to the originating node when the body finishes. Our semantics require checks to ensure that DataCenter is willing to execute the code; that Alice and Bob are both willing to send their schedules to DataCenter; and that Alice and Bob consent to send the available time and resulting calendar modifications to each other. The **hyp** block ensures that the two calendar updates occur atomically.

To design the **with** and **at** constructs, we define *private data* as any data stored on a user's machine that is not marked as public. Users agree to disclose private data to other users via a process called *endorsement* (as in information-flow systems [14]), which is an application-defined procedure that may involve user interaction. Our design ensures that developers control how and when data is endorsed while users retain control over the specific policy for endorsement.

### 4.1 at: Placing Code Execution

Opal's **at** construct lets a program request to run code on another user's machine. Despite the physical location, however, **at** does not grant access to the data on the remote machine. Formally, a block of the form **at** $U$ { s }, where $U$ is a user from the predefined set of users in the system, requests to execute statement s on the node owned by $U$. If $U$ accepts, then the system serializes s along with its closure environment and sends it to $U$ for execution. The remote node responds with the code's results: i.e., variables that are read outside of the **at** block.

While the **at** construct is useful on its own, its primary use within Opal is to allow for *hypothetical* multi-user coordination. Our design therefore extends hypothetical worlds to work across multiple systems controlled by different users. The idea is to let programs enter a hypothetical state that encompasses the entire multi-machine system and choose whether to discard or commit its changes atomically. For example, a calendaring application might make hypothetical rearrangements on multiple users' calendars to make room to schedule a new meeting. Within this hypothetical world, all users observe the same global state. When the hypothetical world commits, it sends simultaneous updates to all users' calendar stores.

In our current prototype, all code placement decisions in Opal are static: the program specifies which node should run each **at** block. In the future, we plan to extend this basic system to automatically negotiate placement decisions based on users' combined preferences. For example, groups of mutually distrustful users may wish to execute all of their shared code on a mutually trusted server, whereas users with strong security requirements may be unwilling to execute code that requires their input anywhere but on their own machines. The placement policy could be decided dynamically based on the results of disclosure requests.

## 4.2 `with`: Requesting Data Access

A second construct, **with**, controls whether data from other users can be disclosed to the machine hosting the computation. A statement of the form **with** $U$ { s }, performs two core functions: it places the user $U$'s data in scope for the statement s, and it provides a scope for which $U$ may endorse the use of its data.

As an example, consider the execution from Bob's perspective when Alice invokes the code in Listing 1. The code accesses `Bob.calendar` while the code is executing **at** `DataCenter`, and then the `fitness` result is sent to code that executes **at** `Alice`. Furthermore, the value of `time` flows to the calendar updates for both users. Therefore, Bob must approve the disclosure to Alice of the values `fitness` and `time`, which are the results of `fitness()` and `find_available_time()` calls, respectively. In general, whenever a **with** $U$ block executes, the user $U$ decides whether or not to endorse a disclosure based on these factors:

- Which machine the code is currently **at**. (The entirety of the private information will necessarily be sent to that machine.)
- Explicit transfers of information inside the **with** block. (For example, `time` flows to Alice in the `Alice.calendar.add(event, time)` call.)
- Which machines have control over data produced in the **with** block. (For example, `Bob.fitness()` flows to the `fitness` variable that Alice will later access.)

After data is disclosed to another user, Opal makes no further guarantees about how that user handles the data. Our threat model includes disclosing data to the wrong user, but not malicious handling by trusted users.

*Granularity of endorsement.* Decoupling the **with** and **at** constructs lets Opal programs choose the *granularity* of data disclosure. For example, consider these two code schematics that differ only in the placement of the **with** block:

```
function fine_grained_endorsement() {
  at A {
    at B {
      with C {
        result = process(C.get_data());
      }
    }
    use(result);
  }
}
```

```
function coarse_grained_endorsement() {
  at A {
    at B {
      with C {
        intermediate = C.get_data();
      }
      result = process(intermediate);
    }
    use(result);
  }
}
```

The two functions accomplish the same task with different endorsement strategies. The first function, `fine_grained_endorsement()`, requires user C to endorse the disclosure of `C.get_data()` to user B and `process(C.get_data())` to A. The second function, on the other hand, requires endorsement to disclose `C.get_data()` to both A and B. Generally, the widest **with** block within a given **at** is the one that necessitates the least permissive endorsement, but this is not always the most desirable configuration. For example, if C is willing to share its data with both A and B equally, then it may be possible to perform the entire computation on A and cut out the overhead from B entirely. In our current prototype, Opal programs precisely specify the endorsement strategy by nesting **with** and **at** blocks. In future work, we plan to exploit program synthesis to automatically find the most permissive placement of **with** blocks that users will consent to endorse.

The `with` block addresses two purposes: it provides users with explanations of which private data will be shared with whom; and it gives developers easy but explicit access to that data. By restricting endorsement to the scope of a small block of code where output is clearly defined, we allow for useful insights from information flow analysis to explain where data will be used. Also, by only allowing remote data to be incorporated within a `with` block, we force developers to acknowledge their use of remote data, and be conscientious of exactly what they are requesting to be endorsed.

*Endorsement interface.* Opal's `with` construct specifies which endorsements are necessary to execute a program, but it does not specify how those endorsements must occur. Instead, Opal exposes an interface that calls a custom procedure to make an endorsement decision. A conservative implementation, for example, might open a dialog for a user every time the Opal runtime requests endorsement; a smarter engine may remember what a specific user has endorsed in the past and not repeat questions. This approach lets the Opal semantics remain maximally conservative with privacy and endorsement and leaves questions about what may be considered overzealous protection to the application—or even up to user preference.

## 5 OPEN QUESTIONS AND OPPORTUNITIES

This paper describes three sets of abstractions in Opal for three core challenges in designing systems that intelligently take actions and adapt while interacting with users. These challenges, however, are only a sampling of the pitfalls in engineering AI-based systems that call out for programming language solutions. For example, real intelligent systems should automatically learn from their experiences over time. We plan to build *reinforcement learning* into the Opal runtime system so that programs can adjust feature weights during execution. Opal programs could become more efficient over time by learning to search more effectively and to create better user experiences by learning personalized ranking functions for individual users based on their actions. Incorporating reinforcement learning will require new abstractions for defining reward signals (i.e., how do we know how good a given decision was?) and credit assignment (i.e, which decision resulted in this outcome?). A second category of challenges surrounds the management of *learning state:* the mutable values that determine the current configuration of a model. Managing this state shares the same set of pitfalls as any global, mutable environment: programs must decide how long the state persists, how to group it into separate modules, and how to allow different modules to benefit from shared learning evidence.

While Opal's feature interfaces raise the level of abstraction for communicating data to external ML implementations, there are more opportunities in managing the application–learning interface. For example, the application could dynamically switch between learning algorithms to find the best one for the particular setting. AI interfaces can also provide a richer, more interactive interface with the user to give insight into its operation. For example, the program might explain to the user why it made particular decisions to help the user better understand it, and the user might provide feedback to help it better serve them.

We hope that the initial directions in this paper serve as inspiration for interdisciplinary research that combines insights from the PL, AI, and systems research communities. With the rising importance of intelligent systems and intelligent interfaces, these communities have an opportunity to build the tools and abstractions that will be used to build a new generation of high-profile applications.